\documentclass{aa}
\usepackage{graphicx}
\usepackage{hyperref}
\usepackage{txfonts}
\usepackage{xcolor}
\usepackage{caption}
\usepackage{subcaption}
\usepackage{textcomp}
\usepackage[rightcaption]{sidecap}
\usepackage[mathlines]{linenoaa}
\usepackage{placeins}  
\usepackage{media9}   
\usepackage[normalem]{ulem}

\urldef{\CDAWcatalog}\url{https://cdaw.gsfc.nasa.gov/CME_list/UNIVERSAL_ver2/2024_10/univ2024_10.html}
\urldef{\hxidata}\url{http://aso-s.pmo.ac.cn/sodc/dataArchive.jsp}
\urldef{\stixdatacenter}\url{https://datacenter.stix.i4ds.net/}

\begin{document} 
    \title{Hard X-ray flux rope anchor points in the October 2024 X9 flare}
    
    \titlerunning{Hard X-ray flux rope anchor points in the October 2024 X9-class flare}	
	
    \author{Muriel Zoë Stiefel\inst{1,2}\corrauth{muriel.stiefel@fhnw.ch}
            \and
            Amaia Razquin\inst{3}\corrauth{amaia.razquin-lizarraga@uni-graz.at}
            \and
            Louise Harra\inst{2,4} \email{louise.harra@pmodwrc.ch}
            \and
            Astrid M. Veronig\inst{3,5} \email{astrid.veronig@uni-graz.at}
            \and
            Säm Krucker\inst{1} \email{samuel.krucker@fhnw.ch}
    }

    \institute{
            University of Applied Sciences and Arts Northwestern Switzerland, Bahnhofstrasse 6, 5210 Windisch, Switzerland
            \thanks{The first two authors contributed equally to this work.}
        \and    
            ETH Zürich, IPA, Hönggerberg campus, Wolfgang-Pauli-Str. 27, 8093 Zürich, Switzerland
        \and
            University of Graz, Institute of Physics, Universitätsplatz 5, 8010 Graz, Austria
        \and
            Physikalisch-Meteorologisches Observatorium Davos/World Radiation Center (PMOD/WRC), Dorfstrasse 33, 7260 Davos Dorf, Switzerland
        \and 
            University of Graz, Kanzelh\"ohe Observatory for Solar and Environmental Research, Kanzelh\"ohe 19, 9521 Treffen, Austria
    }
	
    \date{Received August, 2026; accepted November, 2026}
	
    \abstract 
    {In recent studies, signatures of non-thermal hard X-ray (HXR) emission at the anchor point of erupting flux ropes were reported, suggesting that accelerated electrons trapped in the erupting flux ropes can reach the chromosphere and produce Bremsstrahlung emission there. 
    }
    {We investigated the HXR sources observed during the impulsive phase of the X9-class flare SOL2024-10-03T12:18 from AR~13842 and assessed whether any of the HXR sources are associated with the anchor points of the erupting flux rope.
    }  
    {We analysed the evolution and location of the flare ribbon sources and coronal dimmings to diagnose the eruption process by using multi-vantage point data from the Solar Orbiter and Earth-orbiting instruments.
    We combined this with a pre-eruption non-linear force-free magnetic field extrapolation to reconstruct the initial magnetic connectivity and compare it to the observed flare ribbons and dimming locations. 
    }
    {The magnetic field extrapolation shows two flux ropes before the flare onset. Tether-cutting reconnection during the impulsive phase of the flare results in a larger erupting flux rope formed by the reconnection of the two pre-existing flux ropes. The Spectrometer/Telescope for Imaging X-rays on Solar Orbiter observes at least four distinct HXR sources: two flare footpoints and two additional outer sources at the anchor points of the erupting flux rope. 
    }
    {These observations corroborate earlier speculations that non-thermal electrons trapped in the erupting flux rope precipitate along the field lines into the denser chromosphere. 
    }
	
    \authorrunning{Stiefel, M. Z. et al.}
    \maketitle 
    \nolinenumbers

\section{Introduction}\label{sec:introduction}

Solar flares and coronal mass ejections (CMEs) are some of the most energetic events observable in our Solar System and they often occur together in one solar eruptive event. They are driven by a sudden release of free magnetic energy in the solar corona through magnetic reconnection beneath a flux rope. The energy released is converted into heating and acceleration of the particles. Understanding the distribution and motion of the initially accelerated particles is fundamental for flare analysis \citep[e.g.][]{Fletcher_2011, Benz_2017}. One of the most direct ways to trace the energy release is through hard X-ray (HXR) measurements, which provide insights into the hottest plasma and non-thermal electrons in solar flares. Flares observed on the solar disc often yield observations consistent with the standard flare model \citep[e.g.][]{Shibata_1995}. Accelerated electrons precipitate into the denser chromosphere and interact with the surrounding plasma, leading to Bremsstrahlung at flare footpoints \citep[e.g.][]{Saint_Hilaire_2008}. Hot flare loops develop as a result of chromospheric evaporation \citep[e.g.][]{Antonucci_1982}, connecting the HXR footpoints. The chromospheric footpoints are compact sources and are located in a dense plasma region, which leads to strong Bremsstrahlung emission. Occulted flares provide an opportunity to detect fainter HXR emission associated with the erupting CME, when the main flare emission is behind the limb \citep[e.g.][]{Krucker_2008}. Several observations of HXR emission associated with the erupting CME have been made, showing that energetic electrons and hot plasma are also present in the erupting flux rope \citep[e.g.][]{Krucker_2007, Krucker_2008_b, Glesener_2013, Lastufka_2019, Hayes_2024}. In the era of Solar Orbiter \citep[][]{Muller_2020}, multi-viewing perspectives on the Sun have provided a new perspective on occulted flares. For instance, \citet{Krucker_2026} were able to analyse a flare observed from Earth on disc and occulted from the Solar Orbiter perspective. This observation allowed for a more complete picture of the total electron distribution in a solar eruption, combining electrons in the actual flare and electrons trapped in the escaping flux rope. In addition to the faint X-ray source in the escaping CME, recent studies using observations with the Expanded Owens Valley Solar Array (EOVSA) have observed non-thermal microwave emission at the anchor points of the erupting flux rope \citep[][]{Chen_2020}. This was followed by the first report of simultaneous HXR emissions at both the conventional flare footpoints and at the anchor points of an erupting flux rope \citep[][]{Stiefel_2023}. Flare accelerated electrons that are trapped in the erupting flux rope seem to be able to move along the flux rope structure and reach the denser chromosphere, leading to Bremsstrahlung emission at the anchor points, as already suggested by \citet{Shibata_1995}; see their Fig.~1. Early observations already revealed HXR emission associated with erupting flux ropes. \citet{Ji_2003}, for example, reported coronal HXR observations during a failed filament eruption, while \citet{Liu_2009_a} observed HXR sources associated with the legs of the erupting flux rope. \citet{Liu_2009_b} further demonstrated that the dynamics of the erupting flux rope are reflected in the movement of the HXR sources. These observations suggest that energetic electrons are trapped or created within the magnetic structure of an erupting flux rope and emit Bremsstrahlung. Nevertheless, observations of HXR emission at the anchor points of erupting flux ropes with simultaneous flare footpoint emission are rare.

\begin{figure*}[htbp!]
    \centering
    \includegraphics[width=1\linewidth]{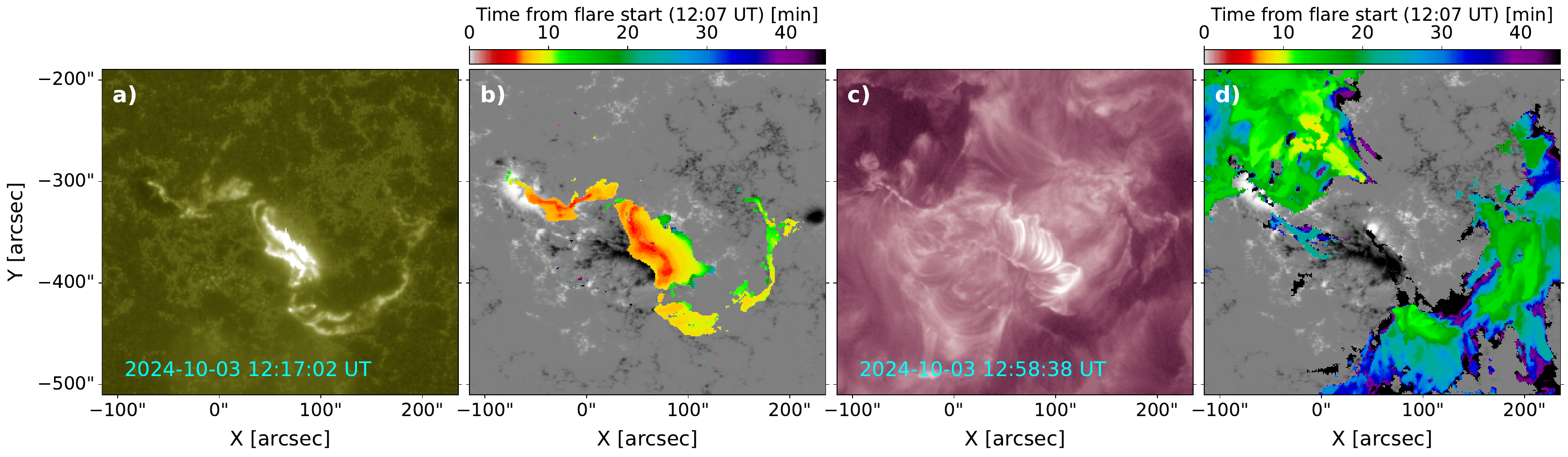}
    \caption{Overview of the evolution of flare ribbons and coronal dimmings as observed in SDO/AIA 1600 and 211~\AA. Panels~a and~b illustrate the flare ribbons: panel~a AIA 1600~\AA~image at 12:17:02~UT, and panel~b ribbon detection mask, showing the temporal evolution of the flare ribbons. Panels~c and~d illustrate the coronal dimmings: panel~c AIA 211~\AA~image at 12:58:38~UT, and panel~d dimming detection mask, showing the temporal evolution of the dimming regions. The associated movie is available online.}
    \label{fig:detection_overview}
\end{figure*}

Identifying erupting flux ropes and their anchor points is inherently challenging, as they are often difficult to directly observe, and the magnetic connectivity evolves throughout the eruption. Before eruptions, flux ropes can be inferred from hot channel structures \citep{zhang2012observation}, such as sigmoids \citep[e.g.][]{mckenzie2008hinode, green2009flux}, filaments \citep{mackay2010physics}, and coronal cavities \citep{wang2010spinning}. Once the eruption begins, coronal dimmings, along with flare ribbons, provide the most straightforward diagnostic for locating the anchor points of the erupting structure. Coronal dimmings are sudden reductions in extreme ultraviolet (EUV) and soft X-ray (SXR) emission caused by plasma evacuation during CME expansion \citep[e.g.][]{hudson1996long, sterling1997yohkoh,  thompson1998soho}, with the most localised and deepest dimmings occurring at the anchor points of erupting flux ropes where plasma depletion is strongest \citep[e.g.][]{attrill2006using, miklenic2011coronal, pan2021preeruption}.
A recent coronal dimming classification framework introduced by \citet{veronig2025coronal} relates observed dimming signatures to the magnetic flux systems involved in the eruption and the processes by which they become involved, providing a physically motivated interpretation of the coronal dimming observations. In particular, the identification of flux rope dimmings and their temporal evolution can provide important insights into the eruption dynamics \citep[e.g.][]{qiu2007magnetic, cheng2016nature,  wang2019evolution, Gou2023complete, wagner2026}. Magnetic field extrapolations, such as non-linear force-free (NLFF) extrapolations, can be used in conjunction to coronal observations to better identify these anchor points and understand the changing magnetic connectivity of the erupting structure \citep[e.g.][]{purkhart2025magnetic, razquin2026magnetic}. 

In the event studied by \citet{Stiefel_2023}, the erupting filament was not directly observable. The analysis relied on multi-wavelength observations to determine that the outer sources were emission from the anchor points of the erupting flux rope. In the present study, we report a similar observation in which five distinct HXR sources are associated with the impulsive phase of the X9 GOES-class flare on the 3 October 2024. Four of the sources are located in the chromosphere, and an additional HXR source is in the corona \citep[the above the loop-top source, reported by][]{Masuda_1994}.
The flare was seen by two HXR instruments: the Spectrometer/Telescope for Imaging X-rays \citep[STIX;][]{Krucker_2020} on Solar Orbiter \citep[][]{Muller_2020} and the Hard X-ray Imager \citep[HXI; ][]{Zhang_2019} on the Advanced Space-based Solar Observatory \citep[ASO-S;][]{Gan_2023}. The event occurred close to the solar equator and was observed on-disc from Earth and at the limb by Solar Orbiter, providing an ideal alignment for multi-wavelength analysis and NLFF extrapolations. In this paper, we present for the first time a comprehensive analysis combining NLFF extrapolations and coronal dimming observations together with HXR observations at anchor points of an erupting flux rope.

\section{Methods}\label{sec:methods}

In this section, we discuss the methods used for the analysis of the data, in particular, flare ribbons and coronal dimming detection, HXR imaging routines, and magnetic field extrapolations. The methods we used are standard methods from previous studies.

\subsection{Flare ribbons and coronal dimming}

We used images from the Atmospheric Imaging Assembly (AIA; \citealt{Lemen2012aia}) on board the Solar Dynamics Observatory (SDO; \citealt{pesnell2012sdo}) to get an overview of the event and detect flare ribbons and coronal dimmings. The images were re-binned to $2048\times2048$ pixels under flux conservation conditions to increase the signal-to-noise ratio, resulting in an effective spatial resolution of 1.2~arcsecs per pixel, and processed with standard Solarsoft IDL software (\texttt{aia\_prep.pro}). We differentially rotated the images to a reference frame 15~min before the flare onset to correct for solar differential rotation. We analysed the event over a 1~hr~15~min period, starting 15~min before the flare. We used full cadence images, which are 12~s for EUV channels (131, 193, 211, and 304~\AA) and 24~s for the 1600~\AA~channel.

For the magnetic context, we used a line of sight (LOS) magnetogram from the SDO/Helioseismic Magnetic Imager (HMI; \citet{Scherrer2012HMI, Schou2012design}) taken at 11:47~UT. We re-binned this image to $2048\times2048$ and processed it with the IDL routine \texttt{hmi\_prep.pro}. We also differentially rotated it to a reference frame 15~min before the flare onset.

We detected flare ribbon regions in AIA 1600~\AA~images with a method similar to that presented by \citet{Kazachenko2017database}. We defined flare ribbon pixels when their intensity exceeded eight times the median image intensity. We cumulatively added each pixel to a ribbon timing mask, where the time of first detection of a pixel was recorded. Figure~\ref{fig:detection_overview} illustrates the evolution of the detection process. Panel~a shows an AIA 1600~\AA~image at 12:17~UT, and panel~b shows the ribbon timing mask over the HMI LOS magnetogram in which the colour represents the time of detection after the flare onset. The accompanying animation shows how the flare ribbons and the detection occur in time.

We detected coronal dimmings in AIA 211~\AA~images using the method developed by \citet{Dissauer2018a}. In this method, a pixel is considered a dimming pixel if the logarithmic base-ratio (log$_\text{10}$) intensity decreases below $-0.19$. We constructed the logarithmic base-ratio images using as a base image the average of three frames taken 15~minutes before the flare onset. Pixels that fulfilled the condition were cumulatively added to a dimming timing mask in which we recorded the first time a pixel was detected as a dimming pixel. Figure~\ref{fig:detection_overview}d shows the timing mask on the HMI LOS magnetogram, where the colour indicates the time of detection after the flare start time. The accompanying animation shows how the detection evolves considering the 211~\AA~observations in panel~c. 

\subsection{Hard X-ray imaging}

In this study, we combined observations of two HXR instruments, STIX \citep[][]{Krucker_2020} on the Solar Orbiter and HXI \citep[][]{Zhang_2019} on the ASO-S. Both instruments provide temporal, spatial, and spectral information on HXR emission in solar flares. STIX observes X-rays in the 4-150 keV energy range, whereas HXI observes between 10-300 keV. For imaging, both instruments use an indirect imaging system by measuring Fourier components, called `visibilities' in HXR imaging. Using these visibilities, images of specific energy ranges can be reconstructed.  

In the following, all times are given in Earth time. Throughout the paper, we corrected the times reported by Solar Orbiter by the additional light-travel time of 350~s from the location of Solar Orbiter (0.299 AU) to 1 AU distance from the Sun. 

\subsubsection{Imaging with Solar Orbiter/STIX}

For the analysis of the STIX observations, we used L1 pixel data from the STIX data centre\footnote{\stixdatacenter} \citep[][]{Xiao_2023} with up to 1~s time resolution in the data. For imaging, we used the standard routines available in the STIX ground-software (GSW) in \texttt{sswidl} \citep[][]{Massa_2023}. We used the reconstruction methods \texttt{mem\_ge} \citep[maximum entropy method;][]{Massa_2020}, CLEAN \citep[][]{Hogbom_1974, Catalano_2026}, and forward fit \citep[][]{Volpara_2022} using the 24 visibilities of STIX with the coarsest grids. We reconstructed the images using all three methods to check the robustness and consistency of the reconstructions \citep[see also the discussion in][]{Stiefel_2023}. In the forward fit method, the number of sources needs to be defined before reconstruction. We used two to five circular Gaussian sources. The number of sources that we decided on was based on the results of the other reconstruction methods and the ability of the method to place the sources reasonably. Note that the flare was outside of the main field-of-view of STIX \citep[e.g.][]{Massa_2023}, such that only the top pixels of STIX were used for the analysis.

To compare STIX images with those from other instruments, we had to shift the images, as the STIX aspect system is not accurate \citep[][]{Warmuth_2020}. The flare discussed in this paper is just at the limb as observed by Solar Orbiter. Therefore, the accurate position of the STIX image is more difficult to determine. This is because the images of the different perspectives cannot simply be reprojected. In this paper, we used the same shift as discussed in \citet{Palumbo_2026}. These values are determined by using HMI difference images and overplotting loops connecting the footpoints. These loops can then be reprojected to the Solar Orbiter viewing angles, and the STIX contours are aligned accordingly. 

Further, for the comparison between observations from the Solar Orbiter perspective and the Earth perspective, we used a LOS. Given a point above the limb from the Solar Orbiter perspective, we determined the 3D coordinates along the LOS between this point and the observer using the \texttt{Astropy} tool \texttt{SkyCoord}. This 3D object can then be overplotted on any map from any viewing perspective using \texttt{SunPy} tools for overplotting \texttt{SkyCoord} coordinates on a \texttt{SunPy} Map. This represents the LOS from the Solar Orbiter perspective.

For the STIX low-energy light curves (< 15 keV), we used the data from the background (BKG) detector of STIX, as the attenuator moved in at the very beginning of the flare. The BKG detector is the only detector on STIX which is not covered by the attenuator and therefore measures unattenuated light curves \citep[for more details, see][]{Stiefel_2025}.

\subsubsection{Imaging with ASO-S/HXI}

\begin{figure*}
    \centering
    \includegraphics[width=1\linewidth]{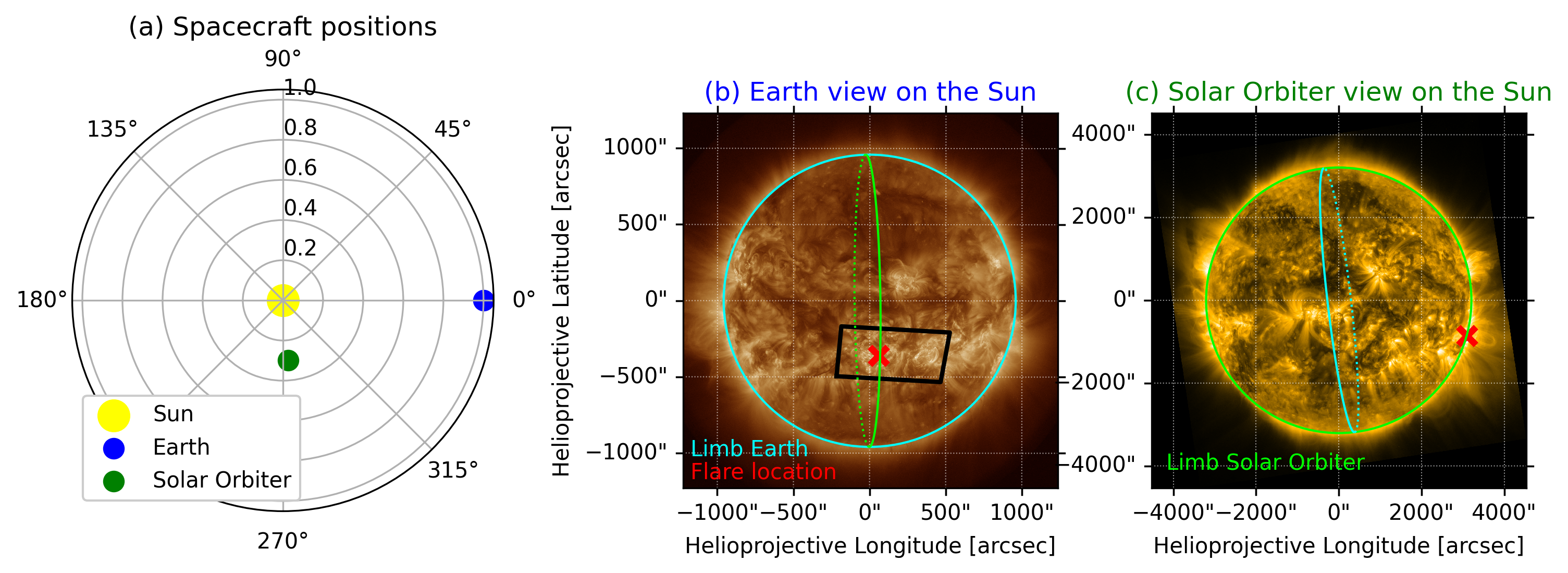}
    \caption{Overview plots for the X9-class flare on the 3 October 2024. Panel~a shows the relative locations of Earth, Sun, and Solar Orbiter. Note that the dot sizes are not to scale. Panels~b and~c show an SDO/AIA 171~\AA~and a Solar Orbiter/Extreme Ultraviolet Imager \citep[][]{Rochus_2020} Full Sun Imager 174~\AA~frame, respectively. The blue limb represents the limb as seen from Earth, whereas the green limb as seen from Solar Orbiter. The red cross marks the flare location. The black box shows the region used for the magnetic field extrapolation as discussed in Section~\ref{sec:nlff}.}
    \label{fig:Overview locations}
\end{figure*}

\begin{figure*}[htbp!]
    \centering
    \includegraphics[width=1\linewidth]{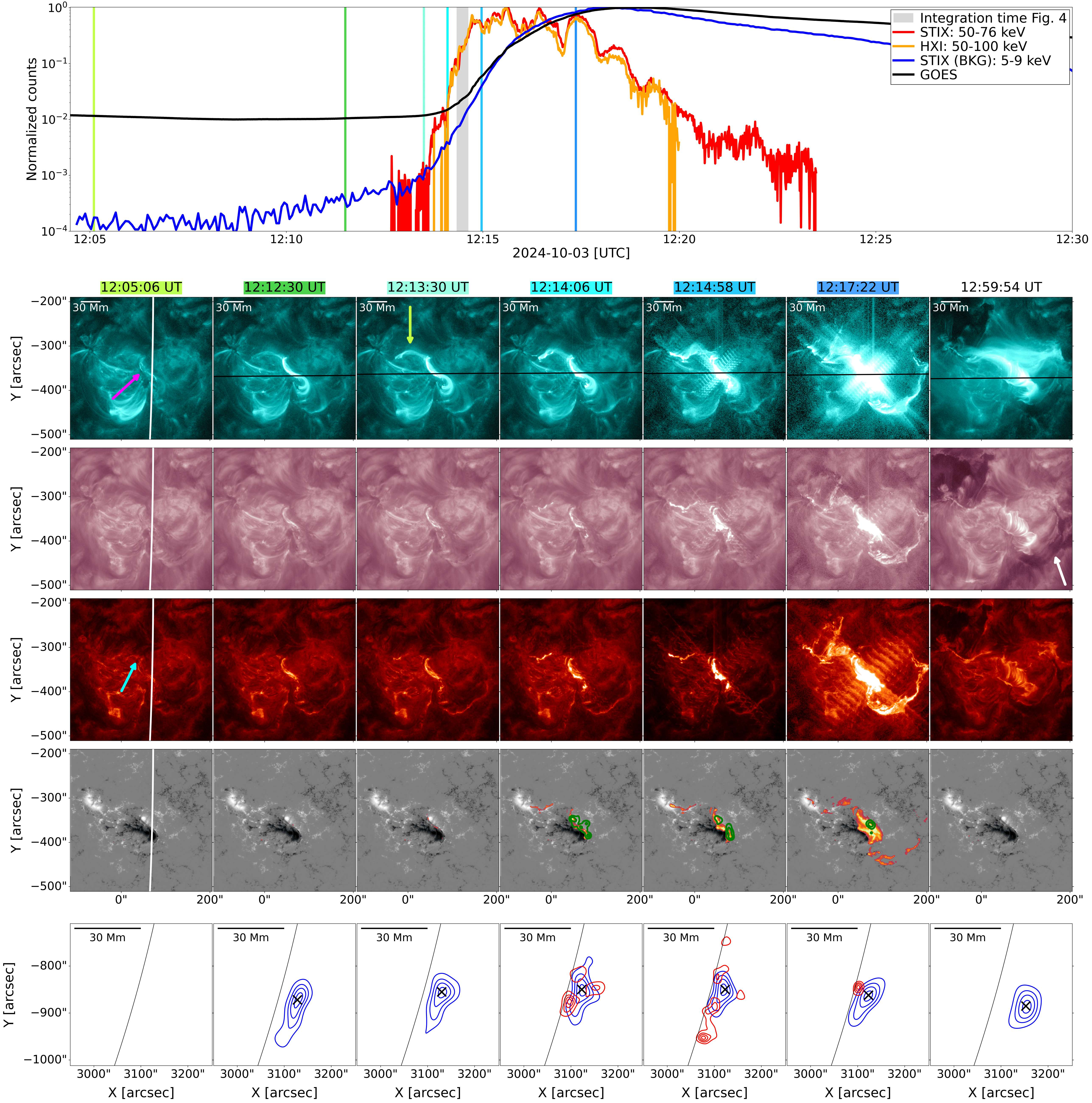}
    \caption{Overview of the flare observed by SDO/AIA, ASO-S/HXI, and Solar Orbiter/STIX at selected time steps. In the top panel, the normalised light curves from STIX, HXI, and GOES are shown. The grey highlighted area marks the integration time range used for the image shown in Fig.~\ref{fig:five_sources STIX}. The coloured lines mark the seven time steps shown in the columns below; note that the last time step is not visible in the light curve. The lower set of images, from top to bottom, AIA 131, 211, 304~\AA~images; HMI LOS magnetogram with AIA 1600~\AA~emission above $150$~DN~s$^{-1}$ in red and HXI 40-55 keV non-thermal contours in green overplotted. In the last row STIX contours are shown, in blue thermal and in red non-thermal contours. The white limb plotted in the first column of the AIA images is the limb of the Sun as seen by Solar Orbiter. The magenta arrow marks the hot loop visible before the flare; the light blue arrow marks dark field lines in AIA 304~\AA~that match the later extension of the hot channel, marked by the yellow arrow; and the white arrow shows the arc-shaped dimming region. The contours of STIX in the bottom row are thermal images (6-10 keV in blue for columns two, three, and seven and 15-18 keV in columns four, five, and six), and non-thermal images (32-76 keV; in red). Contour levels for STIX are 30\%, 50\%, 70\%, and 90\% of the maximum value for each image. The black crosses are used as the STIX LOS, shown as a black line in the AIA 131~\AA~images. As a reference, we show the 30~Mm scale in the 131~\AA~AIA and STIX frames. The associated movie is available online.}
    \label{fig:event_overview}
\end{figure*}

We downloaded the HXI data from the ASO-S data archive\footnote{\hxidata}. For imaging, we used the HXI Graphical User Interface (GUI) in IDL \citep[][]{Su_2019}, which provides the standard imaging routines for HXI. For image reconstruction, we used the CLEAN method. Note that with the current implementation of CLEAN in the HXI GSW, the output maps only include the cleaned beams and do not add the residuals. We checked the reconstructed images by making sure that the residuals were not greater than 30\% to ensure image quality. For the HXI images, we applied the same shift as reported in \citet{Palumbo_2026}, of (2,2) arcsec, to match the AIA observations. HXI light curves were extracted from the GUI.

\subsection{Magnetic field extrapolation}\label{sec:nlff}

We performed the NLFF extrapolation using the NF2 method developed by \citet{jarolim2023probing}. This method is based on a physics-informed neural network (PINN; \citealt{raissi2019pinn}) that solves the force-free magnetic field boundary problem using the observed photospheric vector magnetograms as the lower boundary and a potential field approximation for the upper boundary. The solution is constrained by the divergence-free $\nabla\cdot \mathbf{B}=0$ and force-free $\mathbf{J}\times\mathbf{B}=(\nabla\times\mathbf{B})\times\mathbf{B}=\mathbf{0}$ conditions, where $\mathbf{J}$ refers to the magnetic current density. The neural network provides a mesh-free representation of the extrapolation volume by mapping each coordinate $(x, y, z)$ to a modelled magnetic field $\mathbf{B}=(B_x, B_y, B_z)$. 

The input bottom boundary was set to the space weather HMI active region patch \citep[SHARPs;][]{Bobra_2014} associated with AR~13842 at 11:48~UT. We used the SHARP vector magnetograms in cylindrical equal area (CEA) coordinates with their corresponding uncertainty maps. The SHARP maps contain AR~13842 located close to the central meridian, as well as neighbouring ARs to the west. We constrained the NLFF extrapolation volume to incorporate AR~13842 and the neighbouring AR~13843-4, which resulted in the field of view outlined in black in Fig.~\ref{fig:Overview locations}b. We performed the extrapolation up to a height of 210~Mm, resulting in a computational volume of $534\times252\times210$~Mm. 
To assess the quality of the extrapolated magnetic field, we computed the field-strength-normalised mean absolute divergence, $\langle|\nabla\cdot\mathbf{B}|/|\mathbf{B}|\rangle$, as a measure of solenoidality, and the current-weighted angle $\theta_J$ between $\mathbf{J}=\nabla\times\mathbf{B}$ and $\mathbf{B}$, together with the corresponding current-weighted sine metric $\sigma_J$, as measures of the force-free condition. The resulting values of $\langle|\nabla\cdot\mathbf{B}|/|\mathbf{B}|\rangle=5.07\times10^{-4}$, $\theta_J=6.94$, and $\sigma_J=0.12$ indicate low divergence and a high degree of force-freeness, and are comparable to values reported for other NLFF extrapolations (e.g. \citealt{schrijver2006nonlinear, wiegelman2012how, purkhart2025magnetic}).
To identify regions of strong magnetic twist, we used twist-number maps computed with FastQSL \citep{zhang2022fast}, which is available within the NF2 suite. 

We displayed the results using open-source ParaView\footnote{\url{https://www.paraview.org}} visualisation software. To compare the EUV, SXR, and HXR observations with the Cartesian extrapolation volume, we reprojected the observational data to the CEA projection using SunPy. We note that small projection effects may arise from this reprojection that affect the coronal structures extending away from the photosphere. Thus, we refrained from doing direct comparisons between observed large coronal structures and specific field lines in the extrapolation.

\section{Results}\label{sec:results}

\subsection{Overview of the flare}

In this study, we focused on the X9-class flare on 3 October 2024 \citep[e.g.][]{matsumoto2025data}. The flare started at 12:08~UT and peaked at 12:18~UT.
As shown in Fig.~\ref{fig:Overview locations}a, Solar Orbiter was observing the Sun from a $\sim$90~deg viewing angle compared to the location of Earth and was located very close to the Sun at 0.299 AU. The flare occurred at the west limb as seen by Solar Orbiter (see Fig.~\ref{fig:Overview locations}c) and close to the disc centre from Earth view (see Fig.~\ref{fig:Overview locations}b). An associated halo CME was observed at 12:48:05 UT by the Solar and Heliospheric Observatory \citep[SOHO;][]{Domingo_1995} Large Angle and Spectrometric Coronagraph Experiment \citep[LASCO;][]{Brueckner_1995} C2 telescope with a linear speed of $\sim$820 km/s, as derived from the CDAW catalogue.\footnote{\CDAWcatalog}

Figure~\ref{fig:event_overview} and the accompanying animation show the evolution of the flare as observed from Earth by SDO/AIA and Solar Orbiter/STIX. AR~13842 consisted of three strong polarity regions. The strongest central polarity was a large negative region, west of which a weaker positive polarity formed the main polarity inversion line (PIL) of the AR. Northeast of the central negative polarity, a second strong positive polarity patch was present, while a more diffuse negative polarity surrounded the region to the southwest.

\begin{figure*}
    \centering
    \includegraphics[width=1\linewidth]{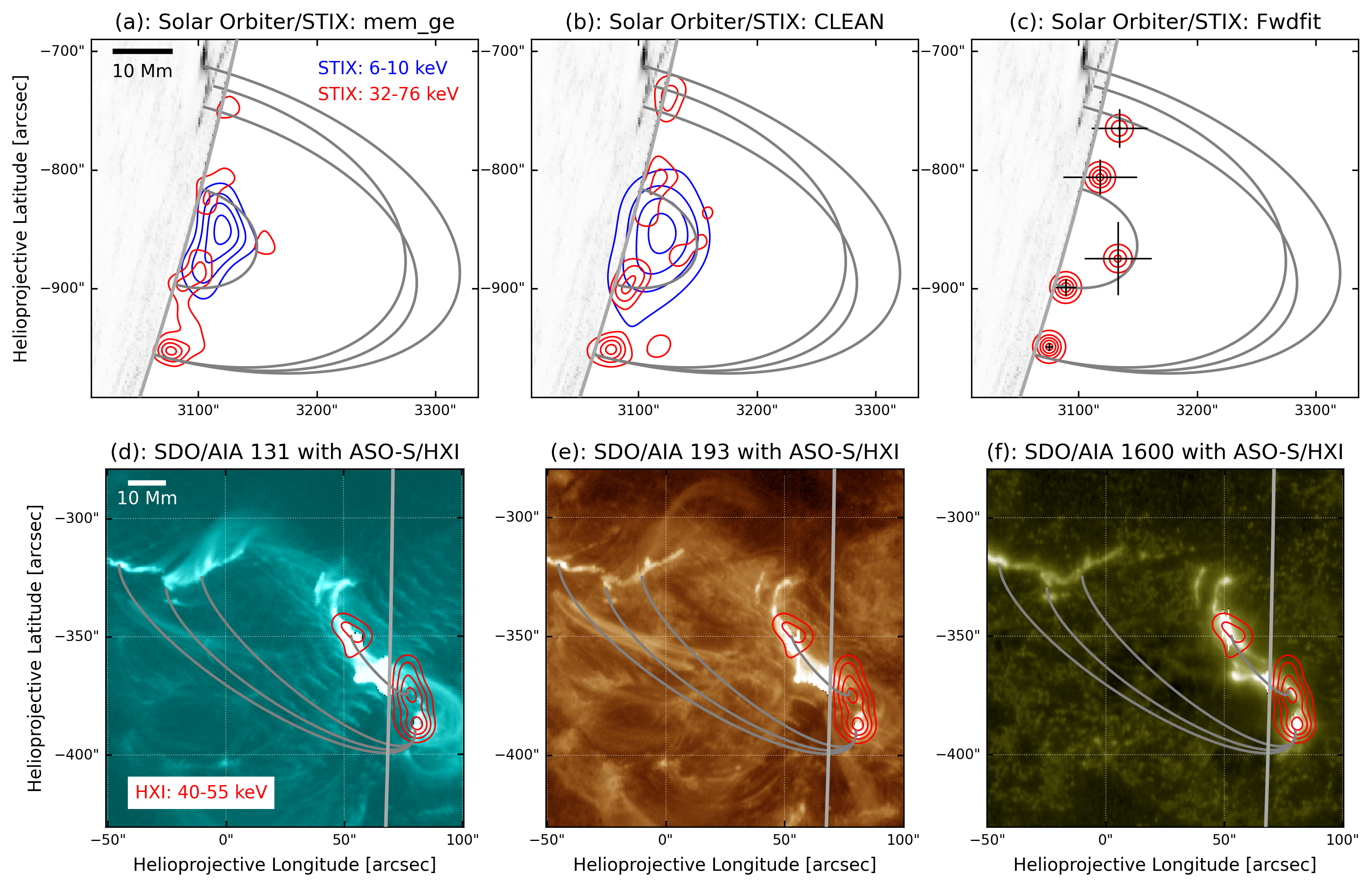}
    \caption{STIX and HXI HXR observations of the flare. Panel~a-c: AIA 1600~\AA~images reprojected to Solar Orbiter view in the background together with the STIX contours. For STIX, the non-thermal images are shown in red and thermal images are in blue. From left to right, we show the results of the reconstruction methods: \texttt{mem\_ge}, CLEAN, and forward fit (using five sources and a fixed source size of 15~\arcsec). The black crosses show the error bars of the position from the forward fit solution. Panel~d-f: AIA images from Earth view, AIA 131, 193, and 1600~\AA~(from left to right). The HXI non-thermal contours are overplotted in red and the limb of Solar Orbiter in grey. The grey loops were added to guide the eye when comparing the two viewing angles. All contour levels are 30\%, 50\%, 70\%, and 90\% of the maximum value. For the forward fit, 10\% levels are shown as well. The associated movie is available online.}
    \label{fig:five_sources STIX}
\end{figure*}

The pre-eruption observations illustrated in the first column show a roughly S-shaped filament-like structure above the main PIL. This structure brightened at the onset of the flare (second column), producing a classic sigmoidal flare configuration. This sigmoid was hot enough to be observed even in HXRs, clearly visible by an increase in the STIX 5-9~keV light curves. STIX images show an elongated structure, which agrees well with the observations of AIA 131~\AA. The black lines in the top row show the STIX LOS of the black cross marked in the bottom panels, projected onto AIA 131~\AA, which illustrates the good alignment between the structure seen by STIX and the S-shaped filament observed by AIA. After 12:13:30~UT (third column, 131~\AA), the hot S-shaped loops extended north-eastwards, marked by the yellow arrow. This is also visible in the STIX images. 
After 12:14:06~UT (fourth row), the impulsive phase of the flare started, showing flare ribbons close to the PIL, and non-thermal footpoints became visible with STIX. The flare continued with the expansion of the ribbons away from the PIL, together with the appearance of additional ribbons at the footpoints of the hot northeastern loops. Around 12:17~UT (sixth column), an additional arc-shaped ribbon region appeared south-west of the active region. Following the main flare impulsive phase, coronal dimming regions developed north-east and south-west of the active region, visible in the 211~\AA~observations at 12:59~UT (see also Fig.~\ref{fig:detection_overview} and its accompanying animation). In particular, a deep and localised arc-shaped dimming region (marked with a white arrow) is clearly visible in the 131, 211, and 304~\AA~channels, occurring in the same location as the earlier arc-shaped flare ribbon.

\subsection{Hard X-ray observations}\label{sec:HXR observations}

In the top panel of Fig.~\ref{fig:event_overview}, the light curves as observed by STIX and HXI in the thermal (low energies) and non-thermal (high energies) emissions are shown during the impulsive phase of the flare. During this time, the non-thermal emission shows a steep increase, including several pulses, typically known as quasi-periodic pulsations \citep[QPP, e.g.][]{Collier_2023}, and decreases again after a couple of minutes. More details on the QPPs for this flare are discussed in \citet{Li_2025}. 

\begin{figure*}
    \centering
    \includegraphics[width=1\linewidth]{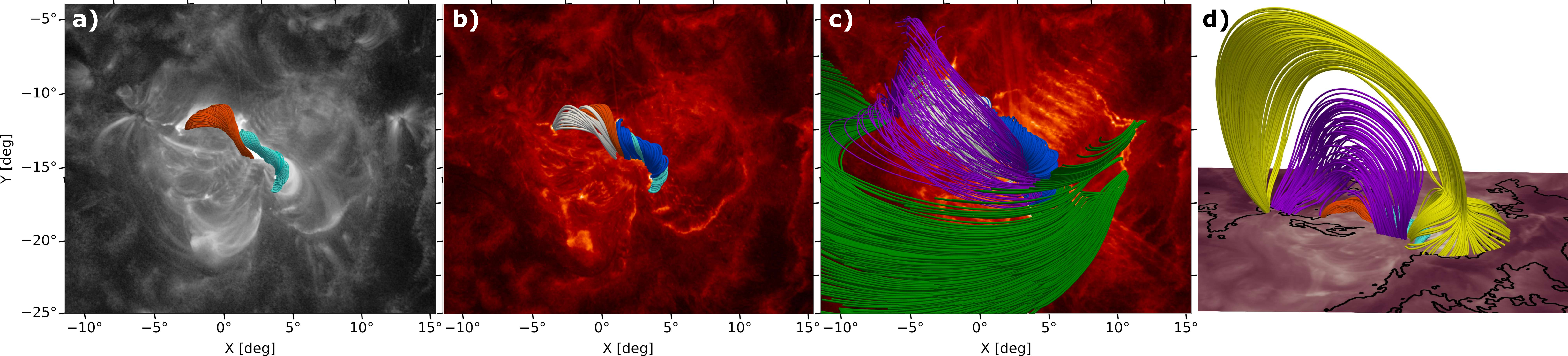}
    \caption{Selected field lines from the NLFF extrapolation overlaid on AIA images. Panel~a: Two pre-existing flux ropes (orange and cyan) over the 131~\AA~channel at 12:14:06~UT. Panel~b and~c: Field lines associated with the flare ribbon regions at 12:14:11 (blue and white) and 12:17:22~UT (blue, white, purple, and green) shown together with the two flux ropes over the 304~\AA~channel at the respective time steps. Panel~d: Flux ropes and selected overlying field lines over a 211~\AA~image at 12:58~\AA; black contours outline the coronal dimming regions. An accompanying animation shows the location of the pre-existing flux ropes with respect to twist maps at different latitudes, where a stronger twist is represented with a darker colour. The associated movie is available online.}
    \label{fig:connectivity}
\end{figure*}

The time range of interest is 12:14:20-12:14:38~UT at the very beginning of the impulsive phase of the flare. In this time step, we observe five distinct non-thermal HXR sources with STIX. The images are shown in Fig.~\ref{fig:five_sources STIX}a-c using the three different reconstruction methods \texttt{mem\_ge}, CLEAN, and forward fit (with five sources as input). HXI observes three non-thermal sources in this time interval, shown in Figure~\ref{fig:five_sources STIX}d-f.
We interpret the five sources observed by STIX in the following way: the two inner footpoints are the standard flare footpoints. They are clearly connected by thermal emission, shown in blue, and visible as well by HXI. The non-thermal source up in the corona is the above-the-loop-top source \citep[][]{Masuda_1994}. The two outer sources are interpreted as additional HXR sources at the anchor points of the erupting flux rope, similar to what was observed in \citet{Stiefel_2023}. HXI also observes a non-thermal HXR source at the southern anchor point. There is no signal visible in the reconstructions of HXI for the most northern source. We propose that this is due to the different viewpoints of the instruments. We discuss the different HXR observations in Section~\ref{sec:HXI-STIX difference} in more detail.

To understand the temporal evolution of the HXR sources, we reconstructed non-thermal images in a 20~s cadence over the impulsive phase of the flare between 12:14~UT and 12:18~UT. The time evolution is available as a movie online, associated with Fig.~\ref{fig:five_sources STIX}. The five sources are only clearly visible for STIX in the time step shown in Fig.~\ref{fig:five_sources STIX}. Before and after, we observed hints that there could still be some emission at the outer source locations, mainly visible in the CLEAN residuals, but it is not consistent over all reconstruction methods. This indicates that the main emission later on is coming from the flare footpoints, and the emission at the anchor points of the flux rope is either lost in the limited dynamic range of the instrument or is no longer present. After 12:16~UT, only emission is visible at the flare footpoints. This is consistent with the observation in \citet{Stiefel_2023} that the outer anchor points are only visible at the very beginning of the impulsive phase. After 12:18~UT, the non-thermal emission decreases drastically, and reconstructions of the footpoints are more difficult. 

Observing five sources at the same time with an indirect imaging system is not straightforward and needs to be treated with care. With STIX, 24 visibilities, meaning 48 numbers, are used to image five sources. With forward-fit, 20 parameters are fitted: 5 sources with four unknowns each, the location $(x,y)$, the size (FWHM), and the flux. To help the system, we fixed the source sizes to 15\arcsec. We report the uncertainties on the position in Fig.~\ref{fig:five_sources STIX}c as errorbars in the image. The coronal source is the most uncertain, which does not matter for the analysis of the flare in this study. The other sources are more robust. Nevertheless, this is at the very edge of the capabilities of what the indirect imaging system can do and is only possible with sufficient statistics. The reconstructed image shown here has roughly 5 times more counts than the event presented in \citet{Stiefel_2023}. As all five sources are consistently observed by the different reconstruction methods, and temporal evolution studies and multi-wavelength analysis support the observations, we believe that there is HXR emission at all five locations. The interpretation of the source morphology and flux is subject to greater uncertainty, and such studies should therefore be approached with caution.

\subsection{Magnetic connectivity}

We identified two main flux ropes from the NLFF extrapolation, shown in orange and cyan in Fig.~\ref{fig:connectivity}a.
We extracted these structures by placing stream tracers in regions of strong magnetic twist within the AR, as illustrated in the movie accompanying Fig.~\ref{fig:connectivity}. The southern (cyan) flux rope closely matches the S-shaped loops that brighten at the onset of the flare around 12:11:30~UT and can also be seen as a filament-like structure in the pre-event observations marked with a magenta arrow in Fig.~\ref{fig:event_overview}. The northern (orange) flux rope can be faintly observed as dark field lines extending north-eastwards from the main PIL in 304~\AA~observations (light blue arrow in Fig.~\ref{fig:event_overview}), which disappear during the eruption. Together, the two flux ropes reproduce the morphology of the hot loop system that appears in AIA 131~\AA~at around 12:13:30~UT. In particular, the northern anchor point of the orange flux rope aligns well with the newly horizontal bright loops observed at the north-east, marked by the yellow arrow in Fig.~\ref{fig:event_overview}. The activation of this bright loops together with the flare ribbons occurring simultaneously on the northeastern positive polarity region and on both sides of the main PIL evidences the reconnection of field lines extending north with field lines of the filament-like structure. Considering the temporal evolution of the event illustrated in Fig.~\ref{fig:event_overview} and the pre-existence of the two flux ropes in the magnetic field extrapolation, the observations suggest that the flux ropes reconnected via tether-cutting reconnection. In consequence, a flux rope anchored in the northern footpoint of the orange flux rope and the southern footpoint of the cyan flux rope was configured and finally erupted. This scenario is also supported by the post eruption arcades observed in AIA 131~\AA~at 12:59~UT, which shows hot loops extending from the northeastern positive polarity region into the main negative polarity region at the centre of the AR.

Figures~\ref{fig:connectivity}b and c show the two flux ropes together with field lines associated with the flare ribbons at two different times, overlaid on AIA 304~\AA~images. In Figure~\ref{fig:connectivity}b, the white and blue field lines were traced from flare ribbons detected up to 12:14:11~UT (see the timing mask in Fig.~\ref{fig:detection_overview}). The white field lines are linked to the flare ribbons in the northeastern positive polarity region, while the blue field lines are connected to the flare ribbons around the main PIL. These field lines correspond to magnetic structures that reconnected during the early stages of the eruption. Some of them appear to be directly involved in the tether-cutting reconnection process, as they closely envelop the two pre-existing flux ropes and may represent the newly formed, larger flux rope resulting from the reconnection.

Figure~\ref{fig:connectivity}c shows in white, blue, magenta, and green field lines associated with the flare ribbons up to 12:17:22~UT, which trace the reconnection as the newly reconnected flux rope expands. During this stage, the flare ribbons propagate in three distinct directions. First, the ribbons spread away from the main PIL; this motion is associated with the blue field lines and likely corresponds to reconnection occurring beneath the rising flux rope. Second, the ribbons extend northwards into the strong positive polarity region. These ribbons are linked to the white and purple field lines, which appear to correspond to strapping magnetic flux; that is, ambient magnetic flux that vaults over and holds down the current-carrying flux. Third, the ribbons expand in an arc shape south-west of the AR. The associated field lines shown in green are predominantly long loops extending over the AR towards the strong positive polarity region in the north-east. This suggests that, as the flux rope expands, progressively larger overlying loops become involved in the eruption and reconnect.

To investigate this further, Fig~\ref{fig:connectivity}d shows two selected magnetic flux systems in purple and yellow over the pre-existing flux ropes. The magnetic configuration is shown from a south-to-north viewing angle with a slight downward inclination over an AIA 211~\AA~image at 12:58~UT. Black contours outline the coronal dimming region (see Fig.~\ref{fig:detection_overview}). The yellow field lines show a fan-spine topology enclosing the southern anchor point of the pre-existing cyan flux rope. The endpoints of the field lines in the fan-spine closely follow both the dimming boundary and the arc-shaped flare ribbons. This indicates that the fan-spine field lines reconnect and expand during the eruption, stretching the surrounding magnetic field and contributing to the south-west dimming region. 
The purple field lines correspond to strapping flux connecting the strong dimming region in the north-east of the AR with the main PIL, where the field lines land almost parallel to the PIL. These field lines are also expected to expand as the erupting flux rope rises. Part of this flux system appears to reconnect and become swept by the flare ribbons, as represented by the purple field lines associated with the flare ribbons shown in Fig.~\ref{fig:connectivity}c.

\subsection{Combining HXR observations with magnetic field extrapolations} 

\begin{figure}
    \centering
    \includegraphics[width=1\linewidth]{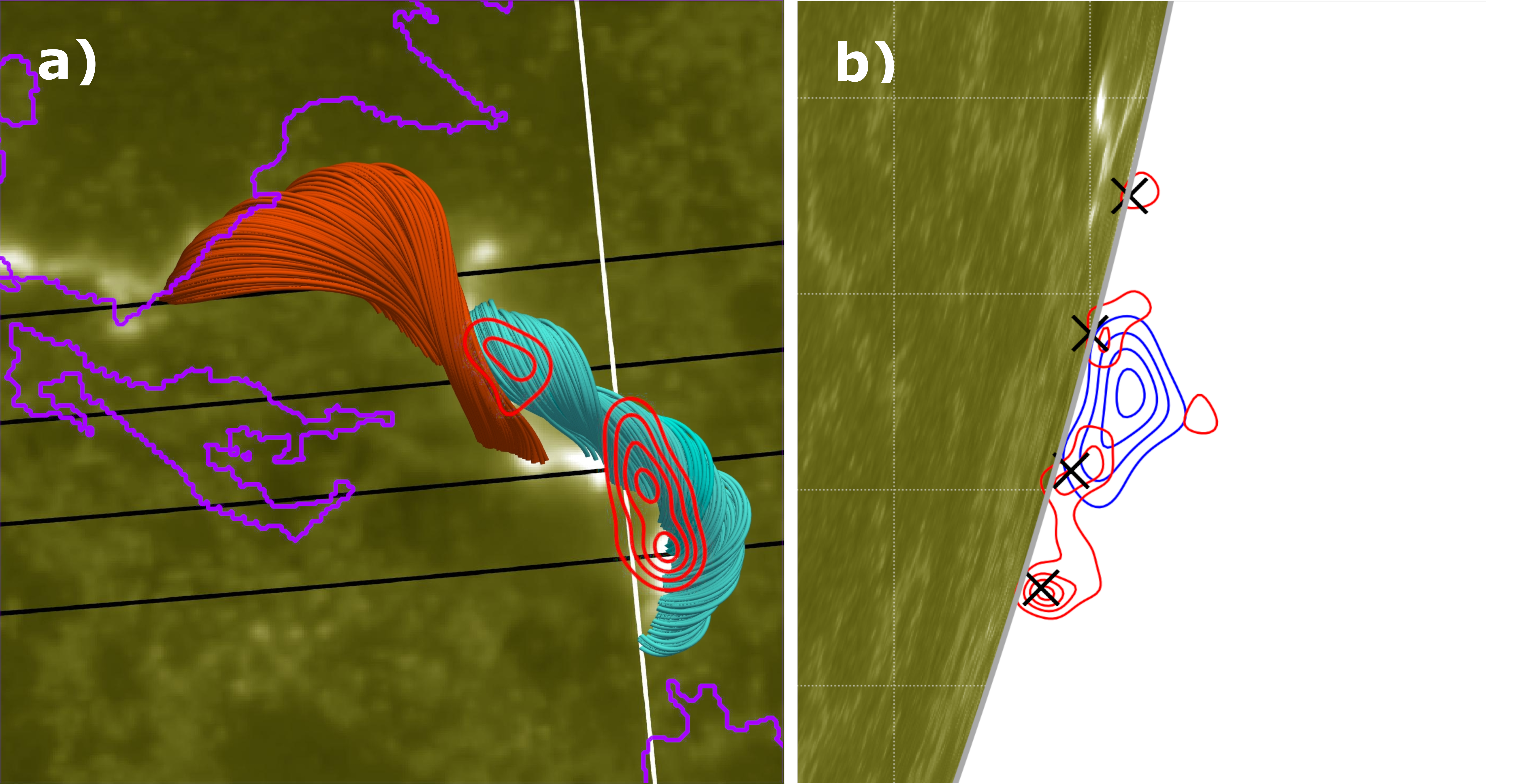}
    \caption{Plots combining magnetic field extrapolations and HXR observations. Panel~a: Flux rope field lines from the NLFF extrapolation (orange and cyan) overlaid on an AIA 1600~\AA~image at 12:14:06~UT. The black lines indicate the LOS line of the non-thermal sources identified by STIX, while red contours mark the HXI non-thermal sources illustrated in Fig.~\ref{fig:five_sources STIX}. The purple contours show the coronal dimming region. The white vertical line represents the solar limb as seen by Solar Orbiter. Panel~b: Reprojected AIA 1600~\AA~image together with the STIX observations. In red are the non-thermal sources, and in blue the thermal emission are shown; see also Fig.~\ref{fig:five_sources STIX}. The black crosses in panel~b show the corresponding positions for the LOS lines shown in panel~a.
    }
    \label{fig:comparison}
\end{figure}

In Section~\ref{sec:HXR observations}, we discussed our interpretation of the individual HXR sources observed by STIX. By combining the HXR observations with the NLFF magnetic field extrapolations and the coronal dimming study, we can support this interpretation. 
Figure~\ref{fig:comparison}a shows the two flux ropes identified in the extrapolation on top of an AIA 1600~\AA~image at 12:14~UT, together with the STIX LOS, HXI source contours, and the coronal dimmings. Figure~\ref{fig:comparison}b shows the STIX observations together with the LOS points in black corresponding to the lines in panel~a. 

The three southernmost STIX sources closely align with the HXI sources. In particular, the southernmost HXR source observed by both instruments is co-spatial with the southern anchor point of the southern flux rope (cyan), whereas the northernmost STIX source coincides with the northern anchor point of the northern flux rope (orange). These two locations correspond to the opposite anchor points of the flux rope formed through the proposed tether-cutting reconnection scenario and conceptually agree with the grey curves given in Fig.~\ref{fig:five_sources STIX}. Thus, this further suggests that the outer HXR sources observed by STIX are associated with non-thermal electrons moving along the erupting flux rope and producing HXR in its anchor points. Comparing the location of the anchor points to the dimming regions shown as a purple contour confirms that these regions are connected with the erupting CME outside of the main flaring region.

The inner sources are consistently observed by both HXI and STIX over the full time range of the impulsive phase of the flare and lie in the regions of the southern footpoint of the orange flux rope and the northern footpoint of the cyan flux rope. In our proposed model, these two regions reconnect through the tether-cutting reconnection scenario. Therefore, the locations of the flare footpoints in HXR trace back to the main reconnection region in the corona where the particles get accelerated and, over the shortest path, reach the dense chromosphere, producing HXR emission. These regions are also observed as flare ribbons in the AIA 1600~\AA~channel as shown in the fourth row of Fig.~\ref{fig:event_overview} and Fig.~\ref{fig:five_sources STIX}f, and discussed in Sect.~\ref{sec:HXR observations}. Comparing the HXI sources with the AIA 1600~\AA~observations, it can be seen that, while the HXI sources appear to involve only the cyan flux rope, the newly reconnected field lines traced by the flare ribbons are anchored Eastwards from the HXI sources. These encompass pre-eruption field lines belonging to the orange flux rope and adjacent sheared field lines. The associated flare ribbons are also well aligned with the LOS of the STIX sources.

\section{Discussion}\label{sec:summary}
We have performed a multi-wavelength and multi-messenger analysis of the X9-class flare on the third October 2024 using SDO, Solar Orbiter, and ASO-S, which compose a two-perspective configuration, with Solar Orbiter viewing the event at a 90~deg angle, on the west limb, very close to the Sun (0.299~AU), while SDO and ASO-S observed it on disc close to the centre meridian. At the time of the event, the AR was composed of a main PIL separating a large, strong negative polarity region from a smaller positive polarity region. In addition, a strong positive polarity region was present to the north-east, while a more extended, weaker negative polarity was to the south-west. The flare began in a classic sigmoid configuration over the main PIL, which developed into a longer structure of bright loops that extended to the positive polarity region north-east of the AR (Fig.~\ref{fig:event_overview}). The accompanying flare ribbons were primarily located on both sides of the main PIL, with additional ribbons developing in northeastern positive polarity region. As the event progressed, an arc-shaped flare ribbon region was formed in the weaker negative polarity region to the northwest of the AR. (Fig.~\ref{fig:detection_overview}b). Similarly, pronounced coronal dimmings developed in the northeast and southwest, with a deep, long-lasting arc-shaped dimming appearing around the arc-shaped ribbon area (Fig.~\ref{fig:detection_overview}d). Notably, during the beginning of the impulsive phase of the event, five HXR sources were identified at the limb by STIX, while HXI could simultaneously observe three of these sources (Fig.~\ref{fig:five_sources STIX}).

To better understand the HXR observations and identify the magnetic structures responsible for the event, we compared them with a pre-eruption NLFF magnetic field extrapolation. The PINN-based NLFF reconstruction method \citep[][]{jarolim2023probing} shows two pre-existing flux ropes (see Fig.~\ref{fig:connectivity}a). The southern flux rope lies above the main PIL, while the northern flux rope connects the negative polarity along the main PIL with the positive polarity region in the northeast. The presence of these flux ropes, together with the initial flare ribbons forming close to the main PIL and the observations of the extended hot channel in AIA 131~\AA~at 12:13~UT (see yellow arrow in Fig.~\ref{fig:event_overview}), suggests that tether-cutting reconnection occurred between the two flux ropes. This reconnection would have produced a larger flux rope rooted at the outer anchor points of the original flux ropes, similar to the magnetic field bundle shown in Fig.~\ref{fig:connectivity}b, which represents pre-eruption field lines associated with the flare ribbons up to 12:14~UT. In addition, coronal dimmings were observed during the impulsive phase of the flare close to the outer end points of the newly reconnected flux rope (Fig.~\ref{fig:comparison}a). This tether-cutting scenario was also observed by \citet{matsumoto2025data}, who performed MHD simulations of this event, where they used an NLFF extrapolation as an initial condition. 

This interpretation closely agrees with the scenario inferred from the HXR observations. The two inner footpoints are persistently observed by HXI and STIX over the impulsive phase of the flare, demonstrating the well-known standard flare model with two non-thermal footpoints and thermal emission connecting the two footpoints. This suggests that the main reconnection location must be above, exactly where the two flux ropes cross each other. These findings support the proposed tether-cutting reconnection scenario and provide convincing observational evidence of previous studies on tether-cutting scenarios \citep[e.g.][]{Xu_2010, Guo_2012}. The outer sources are well aligned with the rooting points of the newly reconnected, larger flux rope, as well as the coronal dimming regions, similar as reported in \citet{Guo_2012}, just that there, no direct HXR signal from the flare itself was observed.  We observe that the outer HXR sources exist for a shorter time period compared to the flare footpoints, consistent with the observations of \citet{Chen_2020} and \citet{Stiefel_2023}. Possible reasons for this can be that the flux rope is expanding in a rapid motion upward, and the electrons that are trapped in the flux rope cannot reach the chromosphere anymore. Another explanation could be that the area where electrons precipitate down along the flux rope field lines increases over time; see also Fig.~\ref{fig:connectivity}d showing the large fan-like area newly connected to the event. This would distribute the emission on to a larger area compared to the very compact flare footpoints and make indirect HXR imagers blind to the emission at the anchor points of the flux rope at later times.

Furthermore, in the MHD simulations by \citet{matsumoto2025data}, the erupting flux rope was anchored in the north-eastern positive polarity region and the negative polarity region where the southern flux rope identified in our extrapolation is rooted. This configuration is consistent with our interpretation and provides additional support for the idea that the observed outer HXR sources are associated with the anchor points of the erupting flux rope. 

\subsection{Coronal dimming interpretation}
The magnetic field extrapolation allows us to interpret the coronal dimming regions within the physics-based categorisation introduced by \citet{veronig2025coronal}. Based on the magnetic connectivity shown in Figs.~\ref{fig:connectivity}c-d, the overall dimming region is largely consistent with a strapping flux dimming. The field lines connecting the south-western negative polarity region to the north-eastern positive polarity region, and arching above the pre-existing flux ropes, were stretched as the newly formed flux rope expanded. The density depletion associated with this stretching produced the resulting strapping flux dimming. 

Flux rope dimmings are expected to form at the anchor points of erupting flux ropes, where plasma depletion is strongest. In the present case, the northern flux rope dimming is difficult to distinguish from the surrounding strapping flux dimming (Fig.~\ref{fig:detection_overview}c), which is caused by the stretching of the strapping magnetic flux overlying the flux rope. In addition, part of the strapping flux shown in purple in Fig.~\ref{fig:connectivity}d seems to be added to the erupting flux rope, since some of them are rooted in the flare ribbon regions at 12:17~UT as shown in purple in Fig.~\ref{fig:connectivity}c. The southern flux rope dimming presents a particularly interesting example of a moving flux rope dimming. These dimmings occur when the erupting flux rope reconnects with surrounding closed magnetic flux, causing the flux-rope footpoint to shift locations and the associated dimming region to migrate. The magnetic extrapolation indicated that the southern footpoint of the southern flux rope is initially rooted near the western edge of the strong negative polarity region. However, this anchor point is embedded within a fan-spine structure. The close correspondence between the fan-spine footprint, the arc-shaped flare ribbons, and the southwestern dimming indicates that these field lines reconnect as the eruption progresses (see Fig.~\ref{fig:detection_overview}b). Consequently, it appears that the erupting flux rope became partially rooted in the fan-spine footprint, which led to a pronounced arc-shaped dimming region within the more widespread southwestern strapping flux dimming region (see white arrow in Fig.~\ref{fig:event_overview}). This interpretation is in line with the MHD simulations by \citet{matsumoto2025data}, which show that the southern footpoint of the erupting flux rope progressively migrated and became anchored in regions associated with the arc-shaped flare ribbons. 

The migration of the southern footpoint of the flux rope may also explain why the northeastern dimming region accurately encloses the northern endpoint of the newly reconnected flux rope, whereas the southwestern dimming is displaced towards the south (Fig.~\ref{fig:comparison}a). As the southern footpoint migrated, the associated flux rope dimming signature evolved more rapidly, while new flare loops developed close to the southern anchor point. In consequence, the initial flux rope dimming became increasingly difficult to distinguish, and only the longer lived flux rope dimming could be detected. This finding further shows the value of the new dimming categorisation framework for identifying and interpreting the observational limitations of coronal dimming signatures (see also \citep{wagner2026}).

\subsection{Different HXR observations between STIX and HXI}\label{sec:HXI-STIX difference}

As shown in Fig.~\ref{fig:five_sources STIX} and discussed in Section~\ref{sec:HXR observations}, HXI only observes three non-thermal sources compared to the five sources observed by STIX. The two flare footpoints and the HXR source at the southern anchor point of the erupting flux rope are jointly observed by STIX and HXI, showing for the first time two observations of different HXR instruments of emission at the anchor points. Due to the top-down view point of HXI on the flare, it is not expected that the above the loop-top source is visible. The location of the loop-top source falls on the location of the flare footpoints.

For the northern flux rope anchor point, we propose that the different viewing angles can explain the difference between the observations of the two instruments. The area in the north where e.g. AIA 1600~\AA~sees emission (Fig.~\ref{fig:five_sources STIX}f) is much larger compared to the southern anchor point from the Earth's perspective. From the Solar Orbiter perspective, the northern area is foreshortened, resulting in a much more comparable size to the southern anchor point. We demonstrate this by using three different loop systems connecting the south with the north in Fig.~\ref{fig:five_sources STIX}. The three loops in the north are anchored far apart on the AIA images from Earth perspective (panels d to f), but they are much closer to each other from the Solar Orbiter perspective (panels b to d). The STIX source is aligned with the ends of the loops in the north. Also, the northern LOS line in Fig.~\ref{fig:comparison}a shows that the STIX emission is integrated over the full ribbon as seen by AIA 1600~\AA~in the north. HXI and STIX must see the same emission in the northern anchor point, but for HXI, the emission is coming from a much larger area such that the signal is lost in the dynamic range of the instrument (indirect HXR imagers typically have at best a dynamic range of $\sim$1:10) and the other three sources are dominating. For STIX, the area over which the emission is distributed is smaller, such that all sources are visible at the same time. This demonstrates how observations are strongly dependent on the viewing angle on the event, especially when using indirect imaging systems due to their limited dynamic range, and could be a significant factor explaining why HXR emission at anchor points of erupting flux ropes is observed so rarely. HXR imaging techniques using direct imaging systems such as the Focusing Optics X-ray Solar Imager \citep[FOXSI; e.g.][]{Krucker_2014,Glesener_2016,Christe_2023} could significantly enlarge our database of similar observations, as the dynamic range is increased compared to indirect imagers. Further simulations are required to investigate the transport of flare-accelerated electrons into the erupting flux rope and their subsequent precipitation into the chromosphere, where they produce X-ray emission.

\section{Conclusion}\label{sec: conclusion}

We investigated the origin of the non-thermal HXR sources observed by STIX during the X9-class flare on the third October 2024. For this, we reconstructed the magnetic connectivity of the event and its evolution using NLFF magnetic field extrapolations and combined this with flare ribbons and coronal dimmings locations. The analysis reveals that the two outer HXR sources are located at the anchor points of an erupting flux rope formed through tether-cutting reconnection between two pre-existing flux ropes, while the inner sources correspond to the standard flare footpoints and coronal source. This observation supports the report by \citet{Stiefel_2023}, where HXR emission at anchor points of an erupting flux rope was detected simultaneously with the flare footpoints, even though the filament was not directly observed and magnetic field extrapolations were not done.
We further observe a moving flux rope dimming alongside arc-shaped flare ribbons which resulted in a dark arc-shaped dimming region. The magnetic field extrapolation shows a fan spine topology with a null-point situated above this arc-shaped dimming, suggesting that one of the anchor points of the flux rope migrated to the foot points of the fan-spine, as the overlying null-point became involved in the reconnection process.

In the present study, combining magnetic field extrapolations with the multi-wavelength observations from quadrature viewpoints provides strong evidence that non-thermal electrons trapped within an erupting flux rope can travel to its anchor points and produce detectable chromospheric HXR emission. Our results highlight the value of combining multi-wavelength observations with magnetic field extrapolations for investigating the physical processes involved in eruptive solar flares.

\begin{acknowledgements}
\em{This paper} is a shared work between the two first authors, Amaia Razquin and Muriel Stiefel, with equal contribution from both authors. The work was conducted using tools provided by SunPy \citep[][]{sunpy_community2020}, Astropy \citep[][]{astropy:2022}, Matplotlib \citep[][]{Hunter_2007}, and NumPy \citep[][]{harris2020array}. M.Z.S. is supported by the Swiss National Science Foundation Grant 10006078. S.K. is supported by the Swiss National Science Foundation Grant 238950. This project has received funding from the European Union's Horizon Europe research and innovation programme under grant agreement No 101134999 (SOLER). The research was sponsored by the DynaSun project and has thus received funding under the Horizon Europe programme of the European Union under grant agreement (no. 101131534). Views and opinions expressed are however those of the author(s) only and do not necessarily reflect those of the European Union and therefore the European Union cannot be held responsible for them. This research was funded in part by the Austrian Science Fund (FWF) 10.55776/I4555 and 10.55776/PIN1569025.
Solar Orbiter is a space mission of international collaboration between ESA and NASA, operated by ESA. The STIX instrument is an international collaboration between Switzerland, Poland, France, Czech Republic, Germany, Austria, Ireland, and Italy. ASO-S mission is supported by the Strategic Priority Research Program on Space Science, the Chinese Academy of Sciences, Grant No. XDA15320000. SDO data are courtesy of NASA/SDO and the AIA. We thank the anonymous referee for the helpful comments that improved the paper.
\end{acknowledgements}

\bibliographystyle{aa}
\bibliography{My_References}

\begin{appendix}
    
\end{appendix}

\end{document}